\documentclass[aps,pre,showpacs,preprint,superscriptaddress]{revtex4}
\usepackage{graphicx}
\usepackage{subfigure}
\usepackage{amsmath}
\usepackage{ulem}
\usepackage{bm}
\usepackage{amssymb}
\usepackage{color}
\newcommand{\ihat}{\bm{\hat{\imath}}}
\newcommand{\jhat}{\bm{\hat{\jmath}}}
\newcommand{\khat}{\bm{\hat{k}}}

\begin{document}

\title{Thomson backscattering in combined uniform magnetic and envelope modulating circularly-polarized laser fields}
\author{Julia Zhu}
\email[Email: ]{juliaz8888@gmail.com}
\affiliation{Phillips Academy, Andover, MA 01810, USA}
\author{Bai-Song Xie}
\email [Corresponding author. Email: ]{bsxie@bnu.edu.cn}
\affiliation{College of Nuclear Science and Technology, Beijing Normal University, Beijing 100875, China}
\affiliation{Beijing Radiation Center, Beijing 100875, China}

\date{\today}
\begin{abstract}
The Thomson backscattering spectra in combined uniform magnetic and cosine-envelope circularly-polarized laser fields are studied in detail. With an introduction of the envelope modulation, the radiation spectra exhibit high complexity attributed to the strong nonlinear interactions. On the other hand, four fundamental laws related to the scale invariance of the radiation spectra are analytically revealed and numerically validated. They are the laws for the radiation energy as the $6$th power of the motion constant exactly, also as the approximate negative $6$th power with respect to the initial axial momentum and laser intensity in a certain of conditions, respectively, and finally an important self-similar law, i.e., when the circular laser frequency, the envelope modulation frequency, and the modified cyclotron frequency are simultaneously increased by a factor, the radiation energy will be increased by the second power of that factor without changing the shape of the spectrum. With the application of these laws, especially the last one, a much higher radiation energy can be obtained and the harmonic at which the maximum radiation occurs can be precisely tuned without changing its amplitude. These findings provide a possible way to advance radiation technology in many fields such as medicine, communications, astrophysics, and security.
\end{abstract}
\maketitle

\section{Introduction}

Due to its rich features and high potential of valuable applications in the radiation fields, scattering produced by electrons moving in laser fields has been studied extensively for almost half a century. As early as 1970, Sarachik and Schappert presented a classical theory of high-intensity Thomson scattering by an electron moving in an arbitrarily intense, elliptically polarized, plane electromagnetic field \cite{Sarachik1970}; and, in 1993, Esarey \textit{et al.} developed a comprehensive theory to describe the nonlinear Thomson scattering of intense laser field from beams and plasmas. They presented examples of possible laser synchrotron source configurations that are capable of generating hard and soft x-rays \cite{Esarey1993}. Salamin and Faisal further extended the study through their multiple publications on relativistic electron scattering in a superintense laser field
\cite{Salamin1996,Salamin1997a,Salamin1997b}. In 2001, Umstadter examined the interactions between plasma electrons and laser light to produce compact laser radiation and caused more studies on improved energy spread \cite{Umstadter2001}. Later, Lau \textit{et al.}, in 2003, presented a classical analysis of Thomson scattering in an intense laser field and first introduced that the scattering spectra is dependent on the amplitude and phase of the electron motion \cite{Lau2003}. He \textit{et al.} also examined electrons accelerated by linearly polarized laser pulses and discovered the equation of the electron's energy gain as a function of the electron's initial position and scattering angle \cite{He2003}.

On the other hand, studies of a charged particle moving in an electromagnetic wave and a constant magnetic field, a setup of which is termed as "autoresonance", were pioneered by several researchers \cite{Magnetic60th}. They found that the particle energy can increase indefinitely at certain conditions. In late 1990s and 2000, Salamin, Faisal, and Keitel presented the spectra of radiation emitted by an electron in a laser field and a uniform magnetic field \cite{Salamin1998,Salamin1999,Salamin2000}. Soon after, Yu \textit{et al.} discovered that backscattered electrons can attain higher energies than forward-scattered electrons and analyzed the electron acceleration caused by linearly polarized laser pulses in a magnetic field and found that radiation occurs at high harmonics of the cyclotron frequency \cite{Yu2000,Yu2002}. In 2005, Gupta \textit{et al.} analyzed electrons in combined oblique magnetic and circularly-polarized laser fields and found the optimal angle of the magnetic field for the highest energy \cite{Gupta2005}. In the following year Singh studied electrons accelerated by a circularly-polarized laser field in an axial magnetic field \cite{Singh2006}.

The autoresonance laser acceleration was further investigated over a wide range of laser and magnetic field parameters by Galow \textit{et al.} in 2013 \cite{Galow2013}. They found that electron energy gains exceeding $100 \rm{GeV}$ are possible under certain conditions. In 2015, Salamin \textit{et al.} \cite{Salamin2015} numerically investigated an electron vacuum autoresonance accelerator scheme which employs circularly polarized terahertz radiation and available magnetic fields and identified the parameters that could make the scheme experimentally feasible.

Recent studies of Thomson backscattering in combined uniform magnetic and polarized laser fields have been focused on the shape of the laser field. For example, in 2016, Fu \textit{et al.} investigated it in the combined fields in which the laser field is circularly polarized \cite{Fu2016}. Through numerical simulations, they found a scale invariance of the Thomson spectrum with respect to the laser intensity and initial axial momentum as scale factors in a high resonant regime \cite{Fu2016}, which can be as a natural but nontrivial extension of previous scaling law for the photon spectral density in \cite{Seipt2011} in the case of the presence of an external uniform magnetic field. Soon after, Jiang \textit{et al.} extended the investigation to the combined fields with an elliptically-polarized laser field. The effects of the initial phase and ellipticity on the backscatter spectra and fundamental frequency were thoroughly analyzed \cite{Jiang2017}.

Although the above publications found a scale invariance of the Thomson spectrum, mechanisms of increasing radiation strength and the tunability of the radiation source are still limited. In this study, we analytically prove and numerically validate that, if we introduce a cosine envelope to the circularly-polarized laser field combined with a uniform magnetic field, a high level of radiation strength and tunability can be obtained. As envisioned, once the cosine envelop is introduced, the radiation spectra exhibit very complicated phenomena. High oscillations appear in the radiation spectra, which are attributed to the strong nonlinear interactions with the interference effect of the electrons motion in the modulated laser pulse field \cite{Brau2004,Umstadter2013}. Obviously this enriches the optical-klystron-like phenomena reported in previous study \cite{Brau2004} and the possible emitted spectral bandwidth controllability \cite{Umstadter2013}. On the other hand, some simple smooth components of the radiation spectrum, named as ARS (Aggregated Radiation Spectra) curves, can be observed and extrapolated, which provides a convenient way to analyze the complicated radiation spectrum.

The advantage of envelop modulating laser field by a cosine enveloping function is that, obviously as an equivalent to the two-color field superposition, it is the highly nonlinear physical process for scattering spectra. Although simple two-color field addition due to the cosine-enveloped field, Thomson backscatter radiation does contain some basic yet very important characteristics. In this study, we uncover and analytically prove four fundamental scaling laws related to the scale invariant of the radiation spectra. The first one states that, for an electron moving in combined uniform magnetic and cosine-enveloped circularly-polarized laser fields, the Thompson backscatter radiation energy is proportional to the $6$th power of the motion constant. The second one states that the radiation spectrum shape is invariant with respect to the axial initial momentum of the electron. Moreover, when the axial initial momentum is much greater than $1$, the radiation energy is proportional to the negative $6$th power of the axial initial momentum. The third one states that, when the laser intensity and the resonant parameter are much greater than $1$, the radiation energy is proportional to the negative $6$th power of the laser intensity and the radiation spectrum shape is invariant with respect to the laser intensity. Lastly, the fourth one states that, when the circular frequency, the cosine-envelope frequency, and the modified cyclotron frequency are simultaneously increased by a factor of $\rho$, the Thomson backscatter radiation energy will be increased by a factor of $\rho^2$ without changing the shape of the spectrum.

The second and third laws are consistent to the numerical findings in the previous studies \cite{Fu2016,Jiang2017}, but the first and fourth laws are new in the present study. The significance of the fourth law can be highlighted by the following. First, the radiation energy can be greatly amplified with a simultaneous increase of the three frequencies of the envelope, the laser field and the cyclotron associated to the external applied uniform magnetic field strength. One example shows that the radiation intensity reaches $0.035$, which is at a high strength we have never seen in all previous studies. Of course, the intensity can be tuned based on the needs by adjusting the scaling factor of these parameters. Second, the harmonic at which the maximum intensity occurs can be precisely tuned by adjusting the circular frequency relative to the enveloping one. This finding can greatly enhance the radiation technology in many fields, such as radiology, astrophysics, and communications.

One of the applications of this study is the production of THz emission. Based on their experiences with existing THz technology, a group of international THz science and technology experts from the fields of medicine, astrophysics, communications, and security repeatedly emphasized the need for tunable, high-power, yet low-cost THz emission \cite{Dhillon2017}. For example, Wallace cited that the fields of dentistry and dermatology could be using THz radiation, but the cost of specialized lasers inhibits the wide purchase of such equipment \cite{Dhillon2017}. With the usage of the cosine-envelope and the subsequent fourth law, high intensity THz radiation can be emitted at a wide range of harmonics. This suggests that, with these findings, lower cost radiation set-ups can be more effective at producing intense radiation, potentially solving this cost dilemma for a variety of fields.

\section{Basic Equations}

We consider the Thomson backscattering by an electron (with mass $m$
and charge $-e$) moving in combined laser and magnetic fields. It is
assumed that the laser field is a circularly-polarized plane wave
with a modulated amplitude dependent upon a cosine function, vector
potential amplitude $A_0$, and laser frequency $\omega=\mu \omega_0$. Note that
$\omega_0$ is just a calibration frequency for the sake of convenience of normalization as below. The laser field propagates in
the positive $z$ direction, and the external uniform magnetic field
$B_0$ is also assumed along the laser propagation direction. The
phase unit of the laser field is denoted as $\eta = \omega_0 \tau -
\textbf{\textit{k}} \cdot \textbf{\textit{R}}$, where $\tau$ is
time, $\textbf{\textit{k}}$ is the laser wave vector, and
$\textbf{\textit{R}}$ is the electron displacement vector defined by
$\textbf{\textit{R}} = X\ihat + Y\jhat + Z\khat$. The combinational
total vector potential of fields can be expressed as
\begin{align}
\ \textbf{A}=A_0 \cos\alpha\eta \left(-\sin\mu\eta\ihat+\cos\mu\eta\jhat\right)+B_0X\jhat.
\end{align}
\textcolor[rgb]{0.00,0.00,0.00}{where $\mu$ represents the laser circular frequency coefficient and
$\alpha$ represents the enveloping coefficient in a sense of calibration by $\omega_0$.} When $\alpha = 0$,
the laser field becomes a circularly-polarized plane wave with a
constant amplitude (i.e. constant enveloping), which has been
studied previously \cite{Fu2016}.

From the vector potential $\textbf{A}$, the corresponding electric
field $\textbf{E}$ and magnetic field $\textbf{B}$ are defined by
the following equations:
\begin{align}
\ \textbf{E}=-\frac{1}{c}\frac{\partial \textbf{A}}{\partial \tau},
\end{align}
and
\begin{align}
\ \textbf{B}=\nabla \times \textbf{A}.
\end{align}
The electron dynamics will be examined by applying the following momentum-energy evolving equations:
\begin{align}
\ \frac{d\textbf{P}}{d\tau} = -e\left(\textbf{E} + \bm{\beta} \times \textbf{B}\right),
\end{align}
and
\begin{align}
\ \frac{d\gamma mc^2}{d\tau}=-ec\bm{\beta}\cdot \textbf{E},
\end{align}
where $\textbf{\textit{P}}$ is the electron relativistic momentum, $\bm{\beta}$ is the electron velocity, and
$\gamma = (1 - \beta ^2 )^{-\frac{1}{2}}$ is the electron relativistic factor. For convenience, we will normalize physical quantities with the following:
\begin{align}
\ t = \omega_0 \tau,\  z = kZ =\frac{\omega_0}{c}Z,\ \textbf{\textit{r}} = \frac{\omega_0}{c}\textbf{\textit{R}},\  \bm{\beta} = \textbf{\textit{v}} = \frac{\textbf{\textit{V}}}{c},\  \textbf{\textit{p}} = \gamma \textbf{\textit{v}} = \frac{\textbf{\textit{P}}}{mc}. \nonumber
\end{align}
Consequently, the phase of the laser field is now denoted as
\begin{align}
\ \eta = t - z. \nonumber
\end{align}

\subsection{The Trajectory Solutions}

From Eqs.(2) and (3), it can be proven that $E_x=B_y$ and
$E_y=-B_x$, where the subscripts $x$ and $y$ respectively represent
the $x$ and $y$ components of $\textbf{E}$ or $\textbf{B}$.
Applying these relationships to Eqs.(4) and (5) yields
\begin{align}
\ \frac{dp_z}{d\tau}=-\frac{e}{mc}\left(\beta_xB_y-\beta_yB_x\right),
\end{align}
and
\begin{align}
\ \frac{d\gamma mc^2}{d\tau}=mc^2\frac{d\gamma}{d\tau}=-ec\left(\beta_xE_x+\beta_yE_y\right)=-ec\left(\beta_xB_y-\beta_yB_x\right).
\end{align}
Hence, $\frac{d\gamma}{d\tau}-\frac{dp_z}{d\tau}=0$. Therefore, the
constant of motion can be obtained as $\zeta =\gamma - p_z =\gamma_0
- p_{z0}$, although the electron moves in a varying-amplitude
circularly-polarized laser field.

By substituting Eq.(1) into Eqs.(2) and (3), and further
substituting the resulting expressions into Eq.(4), we obtain, after
some manipulation, the motion equations of the electron:
\begin{align}
\ \frac{d^2p_x}{d\eta^2} + \omega^2_b p_x = a\Bigl(\left(\omega_b + 2\mu\right) \alpha \sin\alpha\eta \cos\mu\eta + \left(\mu\omega_b + \alpha^2 + \mu^2 \right)\cos\alpha\eta \sin\mu\eta\Bigr) \\
\ \frac{d^2p_y}{d\eta^2} + \omega^2_b p_y = a\Bigl(\left(\omega_b + 2\mu\right) \alpha \sin\alpha\eta \sin\mu\eta - \left(\mu\omega_b + \alpha^2 + \mu^2 \right)\cos\alpha\eta \cos\mu\eta\Bigr)
\end{align}\\
where $a=\frac{eA_0}{mc^2}$ is the normalized laser intensity, $\omega_b=\frac{eB_0}{mc\zeta\omega_0}$ is the modified cyclotron frequency of the electron motion in the combined laser and magnetic fields.

Assuming that, at $t = 0$, $p_x = p_y = 0$, $p_z = p_{z0}$, and
$\eta_{in} = −z_{in}$, the momentum and energy of the electron can
be determined from Eqs.(6), (7), (8), and(9):
\begin{align}
\ p_x = an_1n_2\left(c_1 \cos\omega_b\eta + c_2 \sin\omega_b\eta + q_{xp}\right) = an_1n_2q_x\\
\ p_y = an_1n_2\left(-c_2 \cos\omega_b\eta + c_1 \sin\omega_b\eta + q_{yp}\right) = an_1n_2q_y
\end{align}
\begin{align}
\ p_z &= \frac{(an_1n_2)^2}{2\zeta} \Bigl(c^2_1 + c^2_2 +m^2_1 \sin^2\alpha\eta + m^2_2 \cos^2\alpha\eta +2\left(c_1 q_{xp} - c_2 q_{yp} \right)  \cos\omega_b\eta \\
& + 2\left(c_2 q_{xp} + c_1 q_{yp}\right)  \sin\omega_b\eta\Bigr) + \frac{1}{2\zeta} - \frac{\zeta}{2} \nonumber\\
\ \gamma &= \frac{(an_1n_2)^2}{2\zeta} \Bigl(c^2_1 + c^2_2 +m^2_1 \sin^2\alpha\eta + m^2_2 \cos^2\alpha\eta +2\left(c_1 q_{xp} - c_2 q_{yp} \right)  \cos\omega_b\eta \\
&+ 2\left(c_2 q_{xp} + c_1 q_{yp}\right)  \sin\omega_b\eta\Bigr) + \frac{1}{2\zeta} + \frac{\zeta}{2}, \nonumber
\end{align}
where
\begin{align*}
\ n_1 = \frac{1}{\omega_b-\alpha - \mu},\ n_2 = \frac{1}{\omega_b+\alpha-\mu}, \\
\ c_1= -q_{xp0}\cos\omega_b\eta_{in} - q_{yp0}\sin\omega_b\eta_{in},\\
\ c_2= q_{yp0}\cos\omega_b\eta_{in} - q_{xp0}\sin\omega_b\eta_{in},\\
\ q_{xp}=m_1 \sin\alpha\eta \cos\mu\eta + m_2 \cos\alpha\eta \sin\mu\eta,\\
\ q_{yp}=m_1 \sin\alpha\eta \sin\mu\eta - m_2 \cos\alpha\eta \cos\mu\eta,\\
\ q_{xp0}=m_1 \sin\alpha\eta_{in} \cos\mu\eta_{in} + m_2 \cos\alpha\eta_{in} \sin\mu\eta_{in}, \\
\ q_{yp0}=m_1 \sin\alpha\eta_{in} \sin\mu\eta_{in} - m_2 \cos\alpha\eta_{in} \cos\mu\eta_{in},\\
\ m_1=\omega_b\alpha,\ m_2=\alpha^2-\mu^2+\omega_b\mu.
 \end{align*}\\

Unlike the constant enveloping laser field studied previously \cite{Fu2016}, the cosine-enveloped laser field which is equivalent to two-color field, combined with a uniform magnetic field would exhibit two singular points at $\omega_b =
\alpha + \mu$ and at $\omega_b = -\alpha + \mu$. These two singular points correspond to the exact resonance condition of an electron in combined fields. For this reason, $n_1$ and $n_2$ represent the resonance parameters. Obviously, when $\omega_b$ is close to $\mu+\alpha$, then $n_1 >> 1$; and when $\omega_b$ is close to $\mu-\alpha$, then $n_2 >> 1$. Upon how close one can approach to resonance is that the cyclotron frequency by magnetic field approaches either one of the two color fields. In fact, if there is no amplitude modulating field, the applied magnetic field is as high as $\sim (10^7\zeta-10^8\zeta)$ Gauss for a typical laser field optical wavelength of $\lambda_L=10\rm{\mu m}-1\rm{\mu m}$, where $\zeta$ can reduce the magnetic field significantly if the electron is not at rest initially, for example, $B_0\approx 5\rm{MG}$ when $p_{z0}=10$ for $\lambda_L=1\rm{\mu m}$. On the other hand, by introducing the cosine-function modulation, since $\mu-\alpha$ is usually smaller than $\mu$, the applied magnetic field would be reduced further to some extent, when the resonance is approached. This is also one of the advantages of using a modulated laser field. Because under the exact resonance situations, in which either $1/n_1=0$ or $1/n_2=0$, the problem would be complex and have to be modified by the radiation damping force which can cause a certain of resonance width. In this paper, same as in previous publication \cite{Fu2016} and others, we are not going to study the electron behaviors at the exact resonance cases. Rather, we will focus our study on the electron dynamics and related radiations that are close to or away from the resonance.

The trajectory equations of the electron are obtained via
$\frac{d\textbf{\textit{r}}}{d\eta} =
\frac{\textbf{\textit{p}}}{\zeta}$ as follows:
\begin{align}
\ x(\eta) &= \frac{an_1 n_2}{\zeta} \biggl(\frac{1}{\alpha^2 - \mu^2} \Bigl(\left(- \alpha m_1 + \mu m_2\right) \left(\cos\alpha\eta \cos\mu\eta - \cos \alpha \eta_{in} \cos \mu\eta_{in}\right) \\
& \left. + \left(-\mu m_1 + \alpha m_2 \right) \left(\sin\alpha\eta \sin\mu\eta - \sin\alpha\eta_{in}\sin\mu\eta_{in}\right)\Bigr) + \frac{1}{\omega_b} \Bigl(-c_2 \left(\cos\omega_b\eta - \cos\omega_b\eta_{in}\right) \right. \nonumber\\
& + c_1 \left(\sin\omega_b\eta  - \sin\omega_b\eta_{in}\right)\Bigr)\biggr), \nonumber
\end{align}
\begin{align}
\ y(\eta) &= \frac{an_1 n_2}{\zeta} \biggl(\frac{1}{\alpha^2 - \mu^2}\Bigl((\mu m_1 - \alpha m_2) (\sin\alpha\eta \cos\mu\eta - \sin \alpha \eta_{in} \cos \mu\eta_{in})\\
& - (\alpha m_1 - \mu m_2 )(\cos\alpha\eta \sin\mu\eta - \cos\alpha\eta_{in}\sin\mu\eta_{in})\Bigr) + \frac{1}{\omega_b} \Bigl(-c_1(\cos\omega_b\eta - \cos\omega_b\eta_{in}) \nonumber\\
& + c_2 (-\sin\omega_b\eta + \sin\omega_b\eta_{in})\Bigr)\biggr), \nonumber
\end{align}
\begin{align}
\ z(\eta) &= \frac{(an_1 n_2)^2}{2 \zeta^2} \biggl(c_1 \left(m_1 + m_2\right)\frac{-\cos(\alpha + \mu - \omega_b)\eta + \cos(\alpha + \mu - \omega_b)\eta_{in}}{\alpha + \mu - \omega_b}\\
& - c_1 \left(m_1 - m_2\right)\frac{\cos(\alpha - \mu + \omega_b)\eta - \cos(\alpha - \mu + \omega_b)\eta_{in}}{\alpha - \mu + \omega_b} \nonumber\\
& - c_2 \left(m_1 - m_2 \right)\frac{\sin(\alpha + \mu - \omega_b)\eta - \sin(\alpha + \mu - \omega_b)\eta_{in}}{\alpha + \mu - \omega_b} \nonumber\\
& + c_2 \left(m_1 + m_2 \right)\frac{\sin(\alpha - \mu + \omega_b)\eta - \sin(\alpha - \mu + \omega_b)\eta_{in}}{\alpha - \mu + \omega_b}\nonumber\\
& + \frac{1}{4} \Bigl(2 \left(m^2_1 + m^2_2 \right) \left(\eta - \eta_{in} \right) + \frac{1}{\alpha}   \left(m^2_1 - m^2_2 \right) \left(-\sin 2\alpha\eta + \sin2\alpha\eta_{in} \right)\Bigr) \nonumber\\
& + \left(c^2_1 + c^2_2 \right) \left(\eta - \eta_{in} \right) \biggr) + \frac{1-\zeta^2}{2\zeta^2} \left(\eta - \eta_{in} \right). \nonumber
\end{align}\\

In Eqs. (14) and (15), there seems to be a singular point at $\alpha
= \mu$, but, when $\alpha = \mu$, $m_1=m_2=\omega_b$, allowing
$\alpha - \mu$ to be canceled out from the denominators of both
equations. Additionally, in Eq. (16), there seems to be a singular
point at $\alpha = 0$. However, as $\alpha$ approaches $0$,
$\frac{1}{\alpha} (m^2_1 - m^2_2)(-\sin 2\alpha\eta +
\sin\alpha\eta_{in})$ becomes $(m^2_1 - m^2_2)(-2\eta +
2\eta_{in})$. For this reason, we will separate our equations into
two cases when necessary: for $\alpha = 0$ and for $\alpha \neq 0$.

It is observed from Eqs. (10-15) that the planar trajectory and momentum are periodic with a period of $T=2\pi n$, where $n$ is a smallest integer that makes each of the following terms an integer: $n\alpha$; $n\mu$; $n\omega_b$; $n(\omega_b-\alpha-\mu)$; $n(\omega_b+\alpha-\mu)$, which can be simplified as $n = \max(n_1, n_2)$ when the appropriate parameters are chosen to make either $n_1/n_2$ or $n_2/n_1$ an integer, see the Appendix A for details.
Subsequently,
$\textbf{\textit{p}}(\eta + T) = \textbf{\textit{p}}(\eta)$,
$\textbf{\textit{r}}(\eta + T) = \textbf{\textit{r}}(\eta) +
\textbf{\textit{r}}_0$, $\textbf{\textit{r}}_0 = (0,0,z_0)$ where,
for $\alpha \neq 0$,
\begin{align}
\ z_0 = z(\eta + T) -z(\eta)=\frac{(an_1n_2)^2}{2\zeta^2}T\left(\frac{m_1^2 + m_2^2}{2} +c_1^2+c_2^2\right) +\frac{1-\zeta^2}{2\zeta^2}T,
\end{align}
and for $\alpha = 0$
\begin{align}
\ z_0 = z(\eta + T) -z(\eta)= \frac{(an_1n_2)^2}{2\zeta^2}T\left(m_2^2+c_1^2+c_2^2\right) + \frac{1-\zeta^2}{2\zeta^2}T,
\end{align}
which represents the drift displacement of the electron during one period.
Note that, for the case in which $\alpha = 0$ and $\mu = 1$, it can be proven that the trajectory, momentum, and energy equations are all recovered to the same equations as \cite{Fu2016}.

The above equations will be used to study the Thomson backscattering
spectra in the following.

\subsection{Emission Spectra}

It is well known that the radiation energy emitted per unit solid
angle $d\Omega$ and per unit frequency interval $d\omega$ is given
by
\begin{align}
\ \frac{d^2I}{d\Omega d\omega} = \frac{e^2\omega^2}{4\pi^2c}|\textbf{\textit{n}}\times [\textbf{\textit{n}} \times \textbf{\textit{F}}(\omega)]|^2,
\end{align}
where the dimensionless form of vector $\textbf{\textit{F}}(\omega)$ is
\begin{align}
\ \textbf{\textit{F}}(\omega) = \frac{1}{\zeta} \int_{-\infty}^{\infty} d\eta \textbf{\textit{p}}(\eta) \exp\Bigl(i\omega\bigl(\eta -\textbf{\textit{n}}\cdot \textbf{\textit{r}}(\eta)+ z(\eta)\bigr)\Bigr).
\end{align}

Since the electron is in a helical-type periodic motion, following
the same approach as \cite{Fu2016}, the radiation energy can be
decomposed into a radiation spectrum given by
\begin{align}
\ \frac{d^2I_m}{dtd\omega} = \frac{e^2}{4\pi^2c}\frac{1}{\zeta^2} \left(m\omega_1\right)^2\left(|\textbf{\textit{F}}_{mx}|^2 + |\textbf{\textit{F}}_{my}|^2 \right),
\end{align}
where the $m$-th harmonic amplitude is now
\begin{align}
\ \textbf{\textit{F}}_{mx, my} = \omega_1 an_1n_2 \int_{\eta_{in}}^{\eta_{in}+T} d\eta \textbf{\textit{q}}_{x,y}(\eta) \exp \bigl(i m \left(\eta + 2z \right)\omega_1 \bigr).
\end{align}
In the above equation $\omega_1$ is the fundamental frequency of emitted harmonic spectra given by
\begin{align}
\omega_1 &= \frac{2\pi}{T - \textbf{\textit{n}} \cdot \textbf{\textit{r}}_0 + z_0}.
\end{align}

Note that for the case of Thomson backscatter, unit vector
$\textbf{\textit{n}} = (0,0,-1)$. Using Eqs. (17) or (18),
$\omega_1$ can be expressed as, for $\alpha \neq 0$,
\begin{align}
\ \omega_1 = \frac{\zeta^2}{n + na^2(n_1n_2)^2 \left(\frac{m^2_1 + m^2_2}{2} + c_1^2 + c_2^2\right)}.
\end{align}
and, for $\alpha = 0$,
\begin{align}
\ \omega_1 = \frac{\zeta^2}{n + na^2(n_1n_2)^2 \left(m_2^2 + c_1^2 + c_2^2 \right)}.
\end{align}

\subsection{Fundamental Laws of Thomson Backscatter}

In this section, we will reveal four fundamental scaling laws on the
Thomson Backscatter of an electron moving in combined uniform
magnetic and cosine-enveloped circularly-polarized laser fields.

For simplicity, we will assume $\eta_{in} = 0$ but keep in mind that
all conclusions hold true for any $\eta_{in}$.

(1) Scaling law and scale invariant with respect to the motion
constant of the electron: \textit{For an electron moving in combined
uniform magnetic and cosine-enveloped circularly-polarized laser
fields, the Thomson backscatter radiation energy is proportional to
the $6$th power of the motion constant, and the radiation spectrum
is invariant of the scaling induced by the motion constant.}

Since $\eta_{in} = 0$, we have
\begin{align}
\ n_1 = \frac{1}{\omega_b - (\mu + \alpha)},\ n_2 = \frac{1}{\omega_b - (\mu - \alpha)},\ m_1 = \omega_ba,\ m_2 = \alpha^2-\mu^2+\omega_b\mu,\ c_1 = 0,\ c_2 = -m_2, \nonumber
\end{align}
\begin{align}
\ q_x = -m_2\sin\omega_b\eta + m_1\sin\alpha\eta \cos\mu\eta + m_2\cos\alpha\eta \sin\mu\eta, \\
\ q_y = m_2 \ cos\omega_b\eta + m_1\sin\alpha\eta \sin\mu\eta - m_2\cos\alpha\eta \cos\mu\eta,
\end{align}
\begin{align}
\ z &= \frac{(an_1n_2)^2}{2\zeta^2}\biggl(-m_2(\mu + \alpha) \sin\frac{\eta}{n_1} -m_2(\mu - \alpha)\sin\frac{\eta}{n_2} \\
& + \frac{1}{4}\Bigl(2(m_1^2+m_2^2) \eta + \frac{1}{\alpha} (m_1^2-m_2^2)(-\sin2\alpha\eta)\Bigr) + m_2^2\eta \biggr) + \frac{1-\zeta^2}{2\zeta^2}\eta, \nonumber
\end{align}
and
\begin{align}
\ \textbf{\textit{F}}_{mx, my} = \omega_1 an_1n_2 \int_{0}^{T} d\eta \textbf{\textit{q}}_{x,y}(\eta) \exp \left(i m\omega_e(\eta)\right)
\end{align}
where $\omega_e(\eta)$ is an extended frequency defined by
\begin{align}
\omega_e(\eta) = (\eta + 2z)\omega_1.
\end{align}
Substituting Eqs. (24) and (28) into Eq. (30), it gives
\begin{align}
\ \omega_e(\eta) &= \frac{(an_1n_2)^2}{n + na^2(n_1n_2)^2 \left(\frac{m^2_1 + m^2_2}{2} + m_2^2\right)} \biggl( -m_2(\mu + \alpha) \sin\frac{\eta}{n_1} -m_2(\mu - \alpha)\sin\frac{\eta}{n_2}\\
& + \frac{1}{4\alpha} (m_1^2-m_2^2)(-\sin2\alpha\eta) \biggr) + \frac{\eta}{n}. \nonumber
\end{align}
It is evident that $\omega_e(\eta)$ does not explicitly contain the constant of motion. Moreover, neither $q_x(\eta)$ nor $q_y(\eta)$ contains $\zeta$. The only term in Eq. (29) that has $\zeta$ is $\omega_1$ (see Eq.(24)). Therefore, $\textbf{\textit{F}}_{mx, my}$ is linearly proportional to $\zeta^2$. From Eq. (21) it can be readily proven that the radiation spectrum $\frac{d^2I_m}{dtd\omega}$ is linearly proportional to $\zeta^6$. Since $\zeta$ is not present in the integrands of Eq. (29), the shape of the radiation spectrum is independent of the motion constant.

(2) Scaling law and scale invariant with respect to the axial
initial momentum of the electron: \textit{For an electron moving in
combined uniform magnetic and cosine-enveloped circularly-polarized
laser fields, the Thomson backscatter radiation spectrum shape is
invariant with respect to the axial initial momentum of the
electron. Moreover, when the axial initial momentum is much greater
than $1$, the radiation energy is proportional to the negative $6$th
power of the axial initial momentum.}

The invariant of the radiation spectrum with respect to $p_{z0}$ is
readily proven since it is not included in the integrands of Eq.
(29). By definition, we have $\zeta = \sqrt[]{1+p_{z0}^2} - p_{z0}$.
For $p_{z0}>>1$, we have
\begin{align}
\ \zeta \approx \frac{1}{2p_{z0}}, \nonumber
\end{align}
from which, we can conclude that the radiation spectrum
$\frac{d^2I_m}{dtd\omega}$ is linearly proportional to the negative
$6$th power of the initial axial momentum $p_{z0}$ when $p_{z0}$ is
much greater than $1$.

Note that the same law was numerically established in the previous
study for the constant-enveloped laser field case \cite{Fu2016}.

(3) Scaling law and scale invariant with respect to the laser
intensity: \textit{For an electron moving in combined uniform
magnetic and cosine-enveloped circularly-polarized laser fields,
when the laser intensity and the resonant parameter are much greater
than $1$, the radiation energy is proportional to the negative $6$th
power of the laser intensity and the radiation spectrum shape is
invariant with respect to the laser intensity.}

Under the condition of
\begin{align}
\ a^2(n_1n_2)^2\left( \frac{m_1^2 + m_2^2}{2} + m_2^2 \right) >> 1,
\end{align}
where it is easy to hold for high laser intensity and high resonant
parameter. In this case, from Eq. (31) we have
\begin{align}
\ \omega_e(\eta) &= \frac{1}{n \left( \frac{m_1^2 + m_2^2}{2} + m_2^2 \right)} \biggl( -m_2(\mu + \alpha) \sin\frac{\eta}{n_1} -m_2(\mu - \alpha)\sin\frac{\eta}{n_2} \nonumber\\
& + \frac{1}{4\alpha} (m_1^2-m_2^2)(-\sin2\alpha\eta) \biggr) + \frac{\eta}{n}. \nonumber
\end{align}
Since neither $q_x$ nor $q_y$ contains $a$, the laser intensity $a$
is no longer in the integrands of Eq. (29), which proves the scaling
invariant with respect to $a$ under the given condition of Eq. (32).

In addition, Eq. (29) now reads
\begin{align}
\ \textbf{\textit{F}}_{mx, my} = \frac{\zeta^2}{ann_1n_2 \left( \frac{m_1^2 + m_2^2}{2} + m_2^2 \right)} \int_{0}^{T} d\eta \textbf{\textit{q}}_{x,y}(\eta) \exp \left(i m\omega_e(\eta)\right), \nonumber
\end{align}
which indicates that $\textbf{\textit{F}}_{mx, my} \propto a^{-1}$.

Under the condition of Eq. (32), Eq(24) now simplifies to
\begin{align}
\ \omega_1 = \frac{\zeta^2}{a^2n(n_1n_2)^2 \left( \frac{m_1^2 + m_2^2}{2} + m_2^2 \right)}, \nonumber
\end{align}
it is seen that $\omega_1 \propto a^{-2}$. Using these proportions in Eq. (21), it is easy to check that $\frac{d^2I_m}{dtd\omega} \propto a^{-6}$.

Obviously this scaling law has also been numerically established in
the previous study for the constant-enveloped laser field case
\cite{Fu2016}.

(4) Scaling law and scale invariant with respect to the system's
frequencies: \textit{For an electron moving in combined uniform
magnetic and cosine-enveloped circularly-polarized laser fields,
when the circular frequency coefficient, the cosine-envelope
frequency coefficient, and the modified cyclotron frequency are
simultaneously increased by a factor of $\rho$, the Thomson
backscatter radiation energy will be increased by a factor of
$\rho^2$ without changing the shape of the spectrum.}

Assume that $\alpha$, $\mu$, and $\omega_b$ are changed by a factor
of $\rho$ as shown in the following
\begin{align}
\ \alpha'=\rho\alpha,\ \mu' = \rho\mu,\ \omega_b' = \rho\omega_b \nonumber.
\end{align}
Here we use ($'$) to represent a term after the change. In this
case, we have
\begin{align}
\ n_1' = \frac{n_1}{\rho},\ n_2' = \frac{n_2}{\rho},\ T' = \frac{T}{\rho},\ m_1' = \rho^2m_1,\ m_2' = \rho^2m_2,\ c_1 = 0,\ c_2' = \rho^2c_2,\ \omega_1' = \rho\omega_1. \nonumber
\end{align}
Substituting these expressions into Eqs. (26), (27), and (31) gives
\begin{align}
\ q_x' = \rho^2q_x(\rho\eta), \nonumber\\
\ q_y' = \rho^2q_y(\rho\eta), \nonumber\\
\ \omega_e'(\eta) = \omega_e(\rho\eta),\nonumber
\end{align}
and substituting the above equations into Eq. (29) yields
\begin{align}
\ \textbf{\textit{F}}_{mx, my}' = \frac{\zeta^2an_1n_2}{n+na^2(n_1n_2)^2 \left(\frac{m_1^2+m_2^2}{2} + m_2^2 \right)} \int_0^{\frac{T}{\rho}}d(\rho\eta)\textbf{\textit{q}}_{x,y}(\rho\eta) \exp(im\omega_e(\rho\eta)) = \textbf{\textit{F}}_{mx,my}.\nonumber
\end{align}
Substituting the above equation into Eq. (21) proves that
\begin{align}
\ \frac{d^2I_m}{dtd\omega}' = \rho^2 \frac{d^2I_m}{dtd\omega}.
\end{align}
The significance of this law will be discussed in the following section.

\section{Numerical Results and Analysis}

We first calculate the backscattering spectra of the $m$th order
harmonic radiation with $p_{z0} = 0$, $a=1$, $\alpha = 1$, $\mu =
7$, $\eta_{in}=0$, and $\omega_b = 9$ by evaluating the spectra
using harmonic $m$ from $0$ to $600$ with an $m$-step size of $1$. Note that the presented spectra intensity is normalized by $e^2/(4\pi^2c)$, same as previous study \cite{Fu2016}.
The results are shown in Fig. 1(a). Evidently, as $m$ varies, the backscattering radiation spectra oscillate drastically. However,
Fig. 1(a) also exhibits types of continuous behaviors at a specific step size and initial value of harmonic $m$. For example, by utilizing a step size of $m$ as $7$ and an initial value of $m_{in} = 1$, (which means that we only plot the radiation spectrum at $m =1, 8, 15, 22, ...$), we are able to extrapolate a unique "smooth" curve, as shown in Fig. 1(b).
Additional "smooth" curves can be obtained by varying the initial value of $m_{in}=3$ and $m_{in}=4$, as illustrated in Figs. 1(c) and 1(d). In this study, we name these "smooth" curves as the Aggregated Radiation Spectra curves (ARS curves). The main physical significance and conclusion is that the spectra are highly sensitive to the harmonics in some cases, which would be possible as an ultrashort THz source in the time domain.

To compare with the combined constant-enveloped laser field and a
uniform magnetic field studied in \cite{Fu2016}, we calculate the
spectra with $\alpha = 0$, $p_{z0} = 0$, $a=1$, $\mu = 1$,
$\eta_{in}= -2\pi \left(\left(\frac{na}{\zeta} \right)^2 +
\frac{1}{2\zeta^2} - \frac{1}{2} \right)$, and $\omega_b = 1.1 $
(which results in $n=10$) by varying $m$ from $0$ to $2000$ with a
step size of $1$. The results are shown in Fig. 2(a). Note that
these results are consistent with \cite{Fu2016}. We then calculate
the spectra using the values of $\alpha = 0.1$ and $\omega_b = 1$
(which results in $n=10$) while keeping all other parameters
unchanged, shown in Fig. 2(b). Again, we find that high oscillations
occur at lower harmonics until approximately $m = 600$. Two ARS
curves can be extrapolated using even and odd harmonics. It is
interesting to see that the oscillations are constrained by these
two ARS curves. Fig. 2(c) plots the curve with even harmonics, while
Fig. 2(d) plots the curve with odd harmonics. Fig. 2(c) exhibits a
higher peak at a higher harmonic than the peak in Fig. 2(d), but the
graphs converge at around $m=600$. If we closely examine Fig. 2(a),
we can find that high oscillations do occur at very low-order
harmonics. Therefore, it seems that both the constant-enveloped
field and the cosine-enveloped field have high oscillations at very
low harmonics, but solely in cosine-enveloped fields do high
oscillations appear in a wide range of harmonics along with ARS
curves.

\begin{figure}[ht]
\includegraphics[width=6.5in]{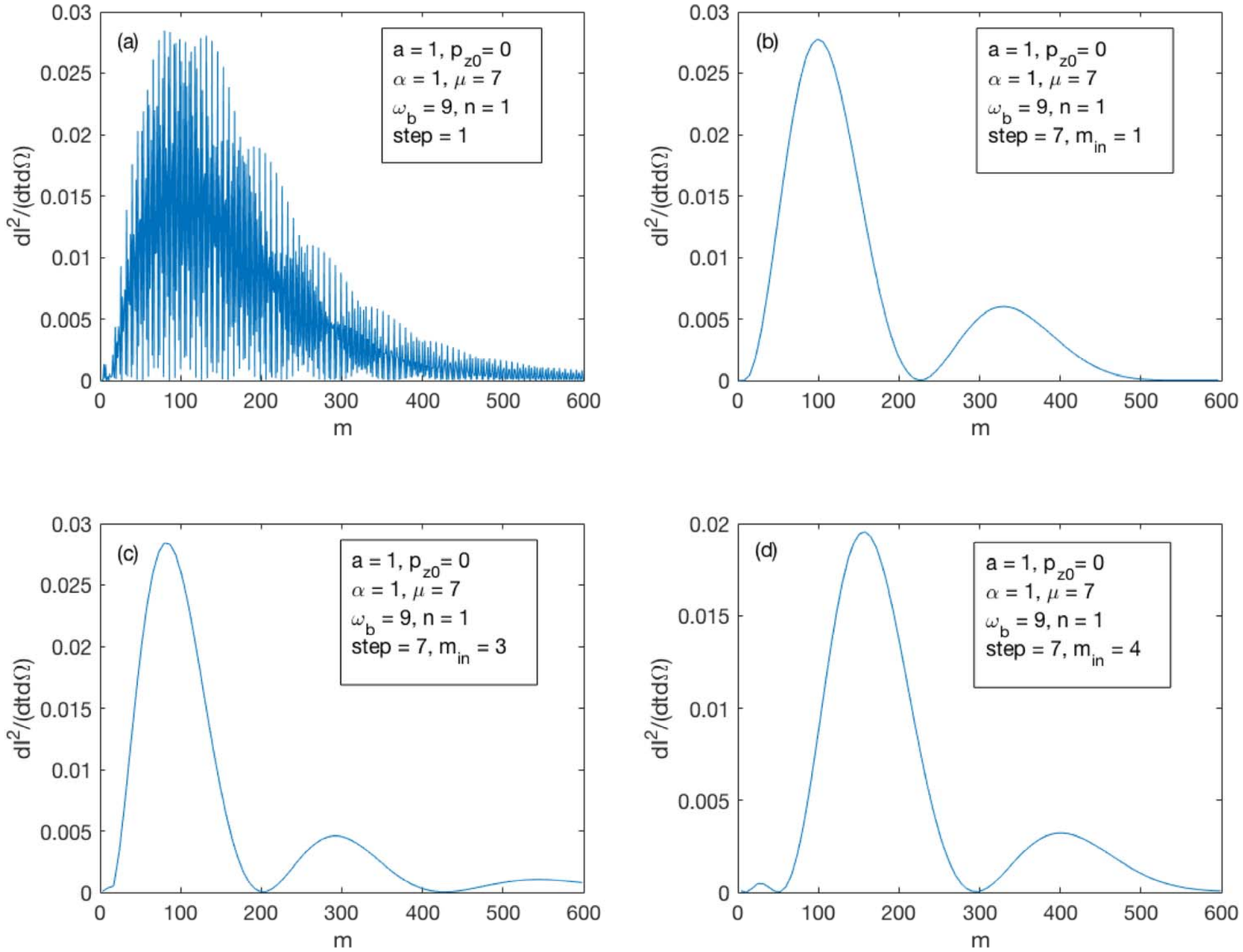}
\caption{The backscatter spectrum of $m$th-order harmonic radiation with different $m_{in}$ and step sizes. (a): $a = 1$, $p_{z0} = 0$, $\alpha = 1$, $\mu = 7$, $\omega_b = 9$ plotting all integer values of $m$ from $0$ to $100$. (b), (c) and (d): step size of $m$ is set as $7$, but with different $m_{in} = 1$, $3$, and $4$, respectively.}
\end{figure}

\begin{figure}[ht]
\includegraphics[width=6.5in]{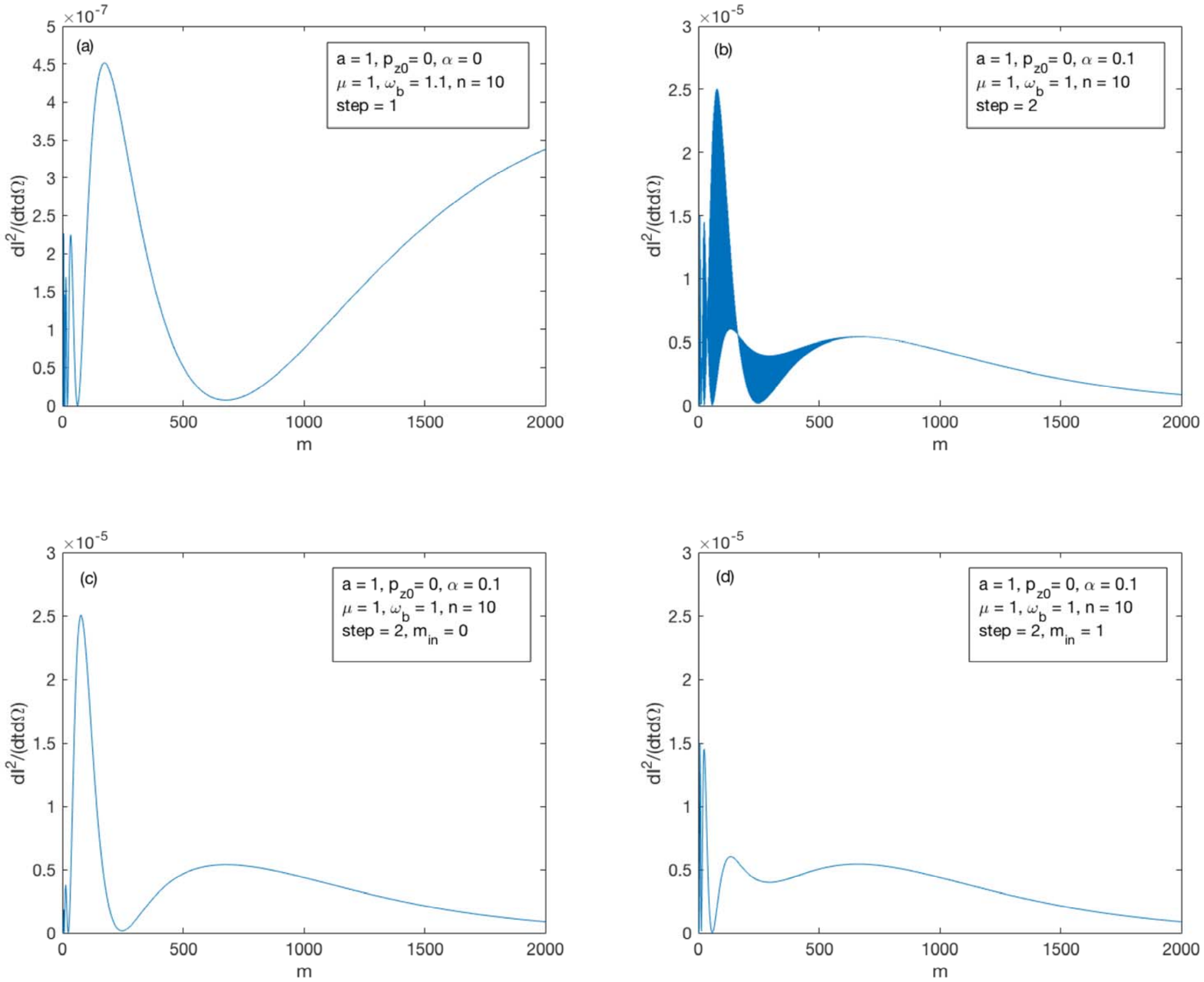}
\caption{The backscatter spectrum of $m$th-order harmonic radiation with different parameters. (a): $a = 1$, $p_{z0} = 0$, $\alpha = 0$, $\mu = 1$, $\omega_b = 1.1$, and $n = 10$. (b): Same as (a) except $\alpha = 0.1$ ($\omega_b$ adjusted accordingly).  (c) and (d): Same as (b) except only even and odd harmonics are plotted, respectively.}
\end{figure}

The physical explanation of the high oscillation phenomenon may be
attributed to the highly nonlinear characteristics of the emitted
spectra brought by the Thomson scattering process (see Eqs. (21) and
(22)). The cosine-enveloping nature of the laser field further
enhances the nonlinear interactions, which leads to the interference
effect of the electrons motion in modulated laser field.

Note that, in order to obtain the same resonant parameter $n=10$,
the modified cyclotron frequency $\omega_b$ needs to be $0.1$ lower
than in the constant enveloping case. Compared to Fig. 2(a), Fig.
2(b) shows that the peak amplitude is much higher but at a lower
harmonic, which indicates that the cosine-envelope is capable of
producing stronger radiation at a lower harmonic. Previous studies
have proven that decreasing the laser intensity increases the
Thomson backscatter radiation \cite{Fu2016,Jiang2017}. In the case
of the laser field discussed in this study, the cosine-envelope
decreases the average laser field intensity, which, in turn,
increases the backscattering. Therefore, besides causing high
oscillations, the enveloped laser field is also effective in
producing high-energy backscatter radiation.

To further enforce the aforementioned conclusions, we study the case
of $p_{z0} = 1$, $a=0.5$, $\mu = 1$, $\eta_{in}= -2\pi
\left(\left(\frac{na}{\zeta}\right)^2+\frac{1}{2\zeta^2}-\frac{1}{2}\right)$,
and $\omega_b = 1.2$ by varying the enveloping coefficient $\alpha$
to $0$, $0.2$, and $0.4$. In order to maintain the same resonant
parameter, $n=5$, $\omega_b$ is modified accordingly. The results
are plotted in Fig. 3. Again, we observe that Fig. 3(a) is
consistent with \cite{Fu2016}, and high oscillations only occur at
very low-order harmonics for both Figs. 3(a) and 3(b). Unlike Fig.
3(a), high-order harmonics dominate the spectrum in Fig. 3(b). Fig.
3(c) provides a closer look at the low-order harmonics of Fig. 3(b).
As $\alpha$ increases, the high oscillations move into higher
harmonic regions as shown in Fig. 3(d), and ARS curves can be
clearly observed. As expected, the radiation amplitude increases
significantly in the case of the enveloped laser field.

\begin{figure}[ht]
\includegraphics[width=6.5in]{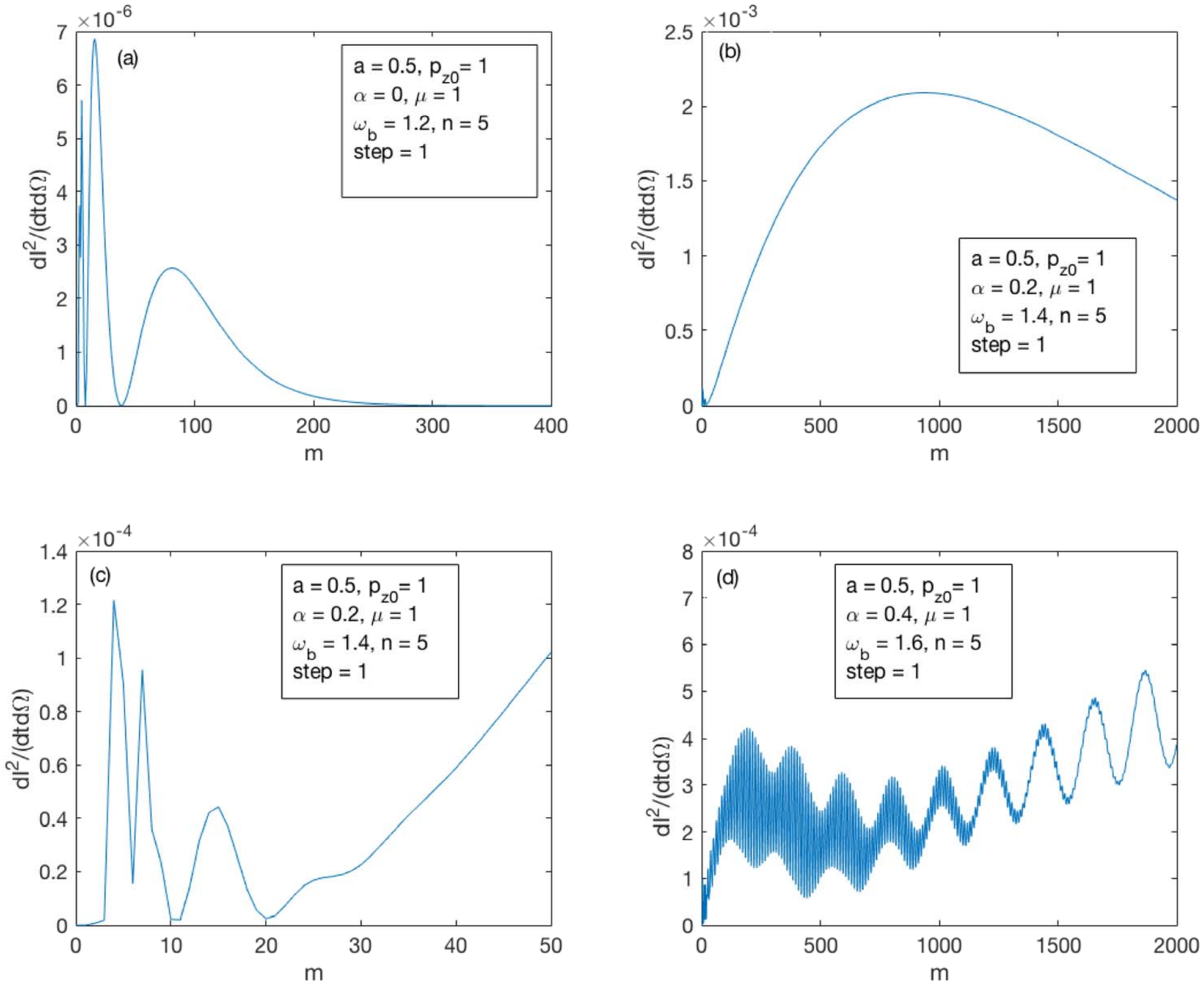}
\caption{The backscatter spectrum of $m$th-order harmonic radiation with different parameters. (a), (b), and (d): $a = 0.5$, $p_{z0} = 1$, $\mu = 1$, and $n = 5$ but $\alpha =0,\ 0.2$, and $0.4$ ($\omega_b$ adjusted accordingly), respectively. (c): Same as (b) except $m$ is only plotted to $50$.}
\end{figure}

In Figs. 4 and 5, we examine the effects of the laser field
intensity $a$ on the backscatter radiation with $\alpha = 0.1$, $\mu
= 0.6$, and $\omega_b = 0.6$ ($n=10$). For simplicity, we
extrapolate an ARS curve using even harmonics. Figs. 4(a)-4(c) plot
$a = 0.5$, $a = 1$, and $a = 2$, respectively, with $p_{z0} = 0$. It
is apparent that, as the laser intensity $a$ increases, the
radiation intensity decreases, which is consistent with the scaling
law of the laser intensity presented in the previous section. Since
the $a$ value is too small to satisfy Eq. (32), the shape of the
curve does not remain unchanged. To satisfy Eq. (32), we increase
$a$ to $20$ and $50$, and the results are plotted in Figs. 5(a) and
5(b), respectively. Evidently, the shapes of the radiation spectra
are almost the same in the two graphs, which numerically proves the
scaling invariant law of the laser intensity.

The effects of the initial momentum $p_{z0}$ are illustrated in
Figs. 4(d)-4(f). Note that Fig. 4(d) is a repeat of Fig. 4(b) for
clarity. From the figures, it is seen that, as $p_{z0}$ increases,
the amplitude of the spectrum decreases, but the shape of the
radiation spectra remains unchanged. As expected, this result is
consistent with the scaling law with respect to the motion constant
and initial axial momentum.

\begin{figure}[htbp]\suppressfloats
\vskip -5cm
\includegraphics[width=17cm]{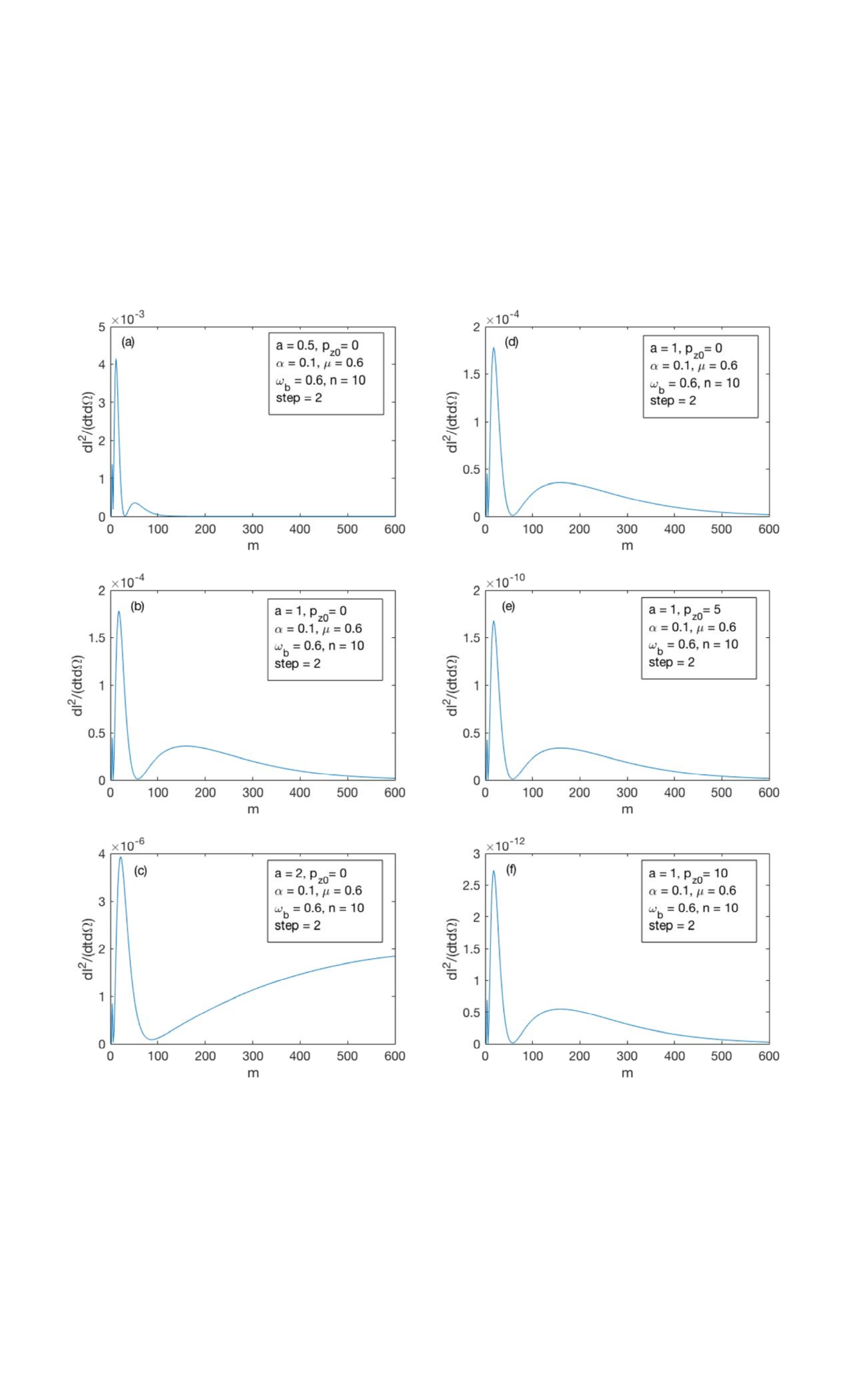}
\vskip -5cm
\caption{The backscatter spectrum of $m$th-order harmonic radiation with different parameters. (a)-(c): $p_{z0} = 0$, $\alpha = 0.1$, $\mu = 0.6$, $\omega_b = 0.6$, $n=10$, and step $= 2$, but $a = 0.5$, $1$, and $2$, respectively. (d)-(f): $a = 1$, $\alpha = 0.1$, $\mu = 0.6$, $\omega_b = 0.6$, $n=10$, and step $= 2$, but $p_{z0} = 0$, $5$, and $10$, respectively.}
\end{figure}

Next, the effects of the circular frequency coefficient $\mu$ on the
spectra are studied in Figs. 6(a) and 6(b), where $a = 1$,
$p_{z0}=0$, $\alpha = 0.1$, $n=10$ ($\omega_b$ adjusted
accordingly), $\eta_{in} = 0$, and $\mu = 0.5$ and $0.7$,
respectively. It is observed that, as $\mu$ increases, the amplitude
decreases and the peaks shift towards higher harmonics. This effect
is further exhibited in Figs. 6(c) and 6(d) where $\alpha$ and $\mu$
and $\omega_b$ are all increased by a factor of ten.

Figs. 6(a) and 6(c) are an example of the fourth scaling law, which
involves $\alpha$, $\mu$, and $\omega_b$ all increasing by a factor
of $10$ ($n$ decreasing by the same factor). The scaling of these
values results in a uniform magnification of the amplitude of the
spectra by $10^2$. Figs. 6(c) and 6(d) demonstrate the same
conclusion. Therefore, our fourth scaling law and scale invariance
of the radiation spectrum is numerically justified.

The findings in Fig. 6 have at least two significance. First, the
radiation energy can be greatly amplified with a simultaneous
increase of $\alpha$, $\mu$, and $\omega_b$. In our example, as
shown in Fig. 6(c), the radiation intensity reaches $0.035$, which
is at a high strength we have never seen in all previous studies. Of
course, the intensity can be tuned based on the needs by adjusting
the scaling factor for $\alpha$, $\mu$, and $\omega_b$. Second, the
harmonic at which the maximum intensity occurs can be precisely
tuned by adjusting the $\mu$ value relative to the $\alpha$ value.
This can greatly enhance the radiation technology in many fields,
such as radiology, astrophysics, and communications.

Because the cosine-enveloped laser field allows for two resonant
parameters, in certain cases, there can exist two values of the
modified cyclotron frequency $\omega_b$ for the same $n$ value.
Figs. 1(a) and 6(d) plot two different $\omega_b$ values with
identical values for all other parameters. When comparing the two
figures, it is seen that the larger $\omega_b$ produces higher peaks
and more ARS curves and moves the peaks to higher harmonics. This
suggests that, since $\omega_b$ is directly proportional to the
magnetic field, a higher magnetic field can both increase the total
energy as well as the complexity of the radiation spectrum.

\begin{figure}[ht]
\vskip -1cm
\includegraphics[width=6.5in]{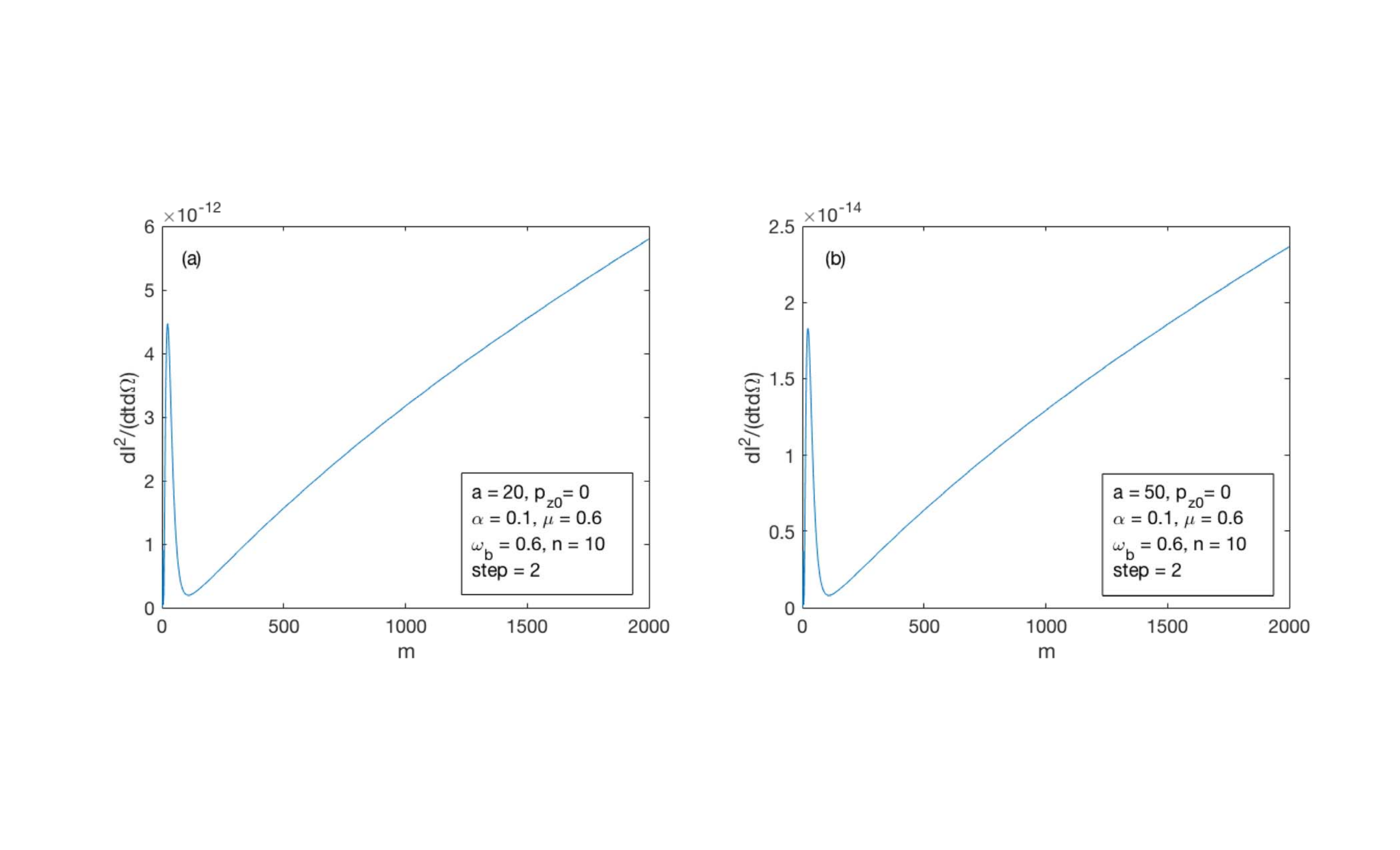}
\vskip -2.5cm
\caption{The backscatter spectrum of $m$th-order harmonic radiation with different parameters. (a) and (b): $p_{z0} = 0$, $\alpha = 0.1$, $\mu = 0.6$, $\omega_b = 0.6$, $n=10$, and step $= 2$, but $a = 20$ and $50$, respectively.}
\end{figure}

\begin{figure}[ht]
\includegraphics[width=6.5in]{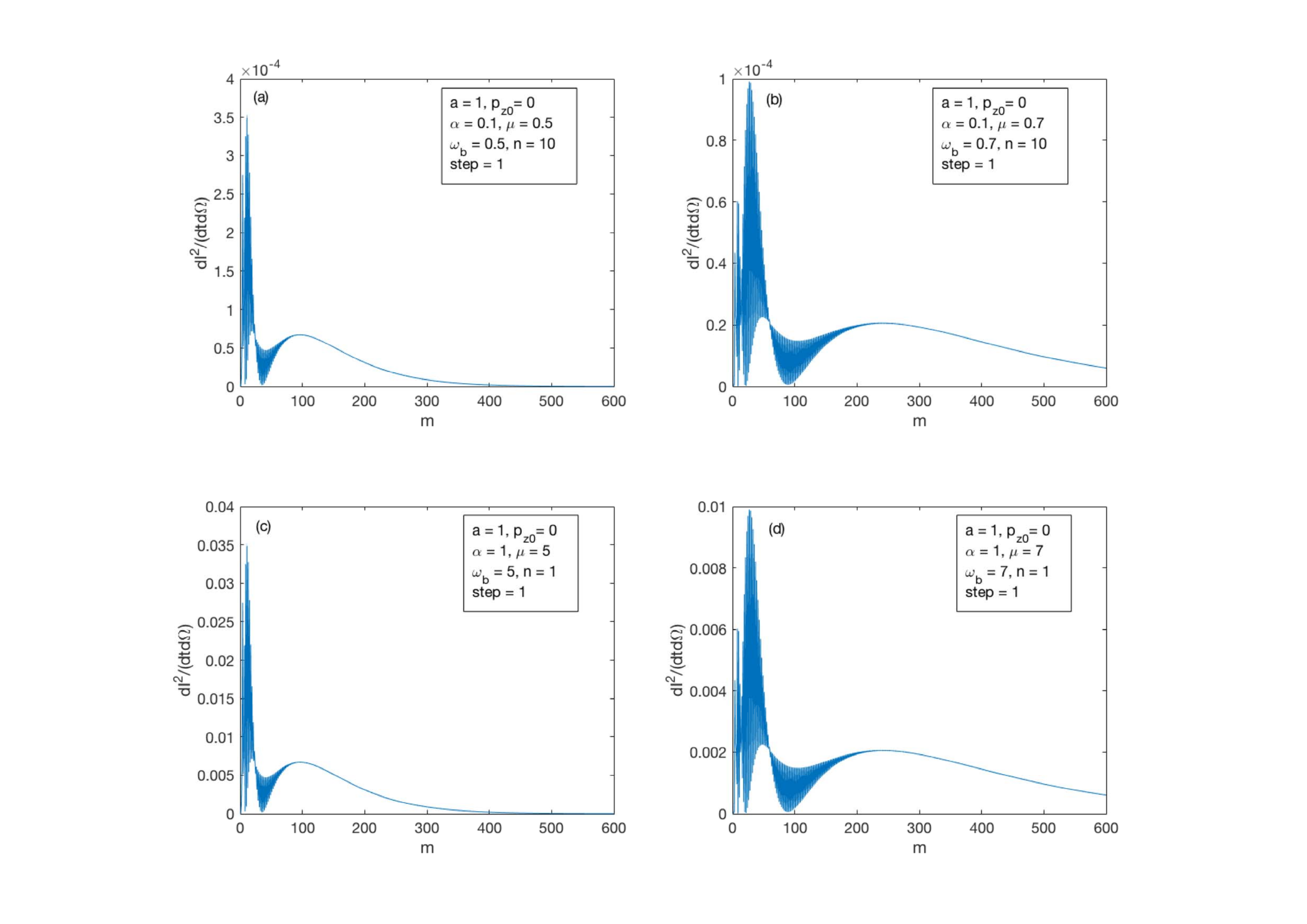}
\caption{The backscatter spectrum of $m$th-order harmonic radiation with different parameters. (a) and (b): $a = 1$, $p_{z0} = 0$, $\alpha = 0.1$, and $n=10$, but $\mu = 0.5$ and $0.7$ ($\omega_b$ changed accordingly), respectively. (c) and (d): Same as (a) and (b) except $\alpha = 1$ and $n=1$,  but $\mu = 5$ and $7$ ($\omega_b$ changed accordingly), respectively.}
\end{figure}

In Fig. 7, we further study the radiation spectra behavior under a
high resonant parameter: when $n = 100$. A constant-enveloped laser
field is compared to a slightly-enveloped laser field in Figs. 7(a)
and 7(b), respectively. Once the cosine-envelope is introduced in
Fig. 7(b), the peak amplitude increases, and the peak frequency
moves to a lower harmonic. These behaviors are consistent with the
conclusions of Figs. 2 and 3. On the other hand, since the
enveloping coefficient $\alpha$ is very small in Fig. 7(b), the
overall behavior is comparable to that of the constant envelope in
Fig. 7(a). Unlike Figs. 2 and 3, however, we do not immediately
observe drastic oscillations with the introduction of the
cosine-envelope.

The absence of high oscillations in Fig. 7(b) deserves further
analysis. We begin by gradually lowering the values of the laser
intensity until $a = 0.55$, graphed in Fig. 7(c), at which high
oscillations begin to form, indicating that $a = 0.55$ is a critical
point. In addition, a strong ARS also forms, which gradually
envelopes the high oscillation peaks. By further lowering $a$ to
$0.53$, the density of the high oscillations increases, and more ARS
curves become noticeable with peaks slightly moving towards
lower-harmonics. Further reductions of $a$ to $0.52$ and $0.5$
exhibit the same overarching ARS curve, but clearer ARS curves with
added complexity appear, resembling those of Fig. 1.

In the final examination of the numerical analysis, we
apply the fourth scaling law to search for and produce high
energy radiation in THz frequencies. Based on the experience from
all the above numerical calculations and observations, we found
that, at $a=1$, $p_{z0} = 0$, $\alpha = 0.01$, $\mu = 0.05$,
$\omega_b = 0.05$, and $n = 100$ and when the typical wavelength
$\lambda = 1 \mu m$ is used, a radiation intensity of about $3
\times 10^{-6}$ can be obtained for harmonics $m=1$ up to $m=9$,
which corresponds to $0.3-3$ THz. These results are plotted in Fig.
8 where only odd harmonics are graphed for clarity. Note that the
radiation intensity found here is $10^3$ higher than what was
obtained in \cite{Jiang2017}. Of course, optimal radiation is
attainable through further tuning.

\begin{figure}[ht]
\includegraphics[width=6.5in]{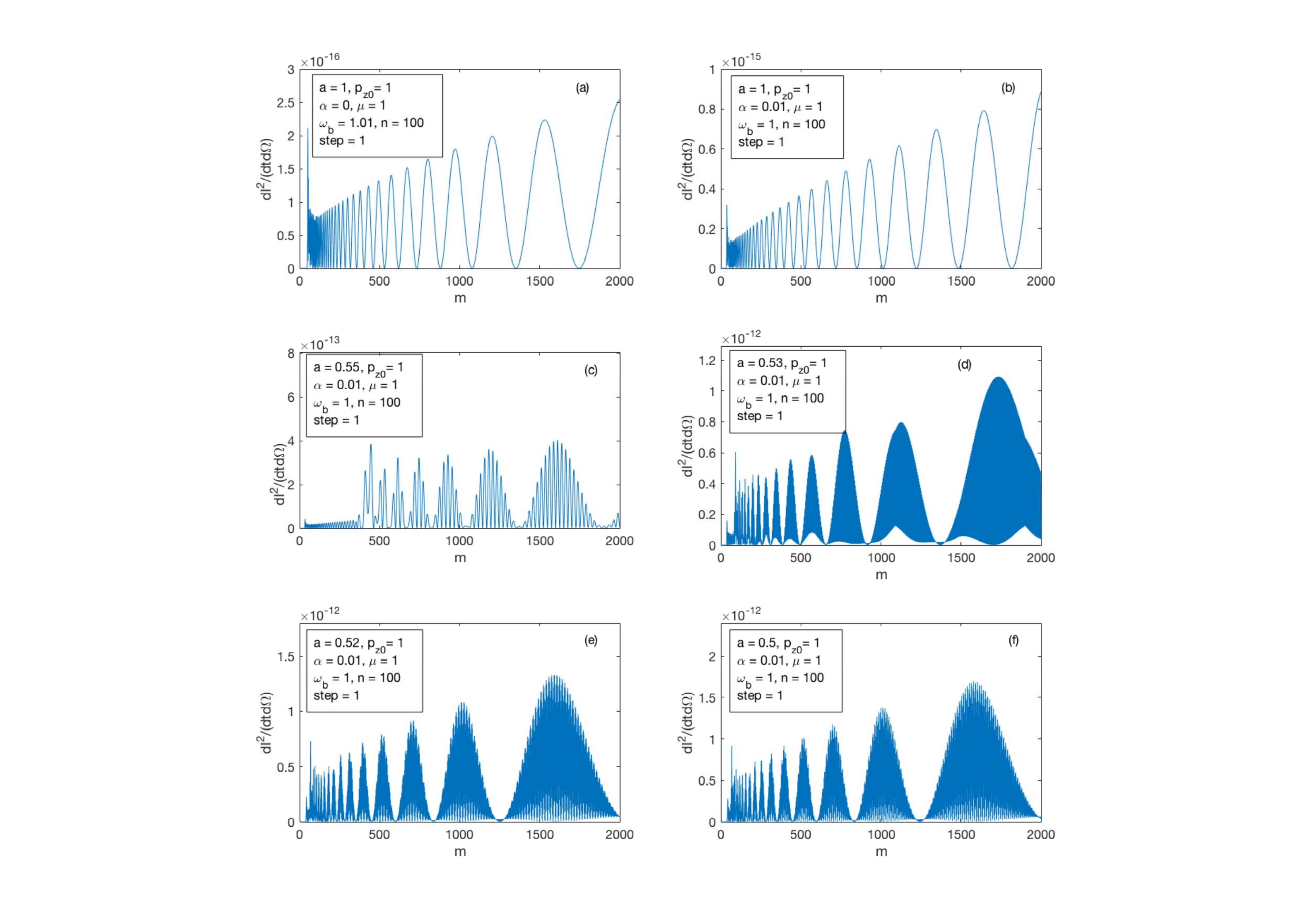}
\caption{The backscatter spectrum of $m$th-order harmonic radiation with different parameters. (a) $a=1$, $p_{z0} = 1$, $\alpha = 0$, $\mu = 1$, $\omega_b = 1.01$, $n = 100$. (b): Same as (a) except $\alpha = 0.01$. (c)-(f): Same as (b) except $a = 0.55$, $0.53$, $0.52$, and $0.5$, respectively}
\end{figure}

\begin{figure}[ht]
\includegraphics[width=6.5in]{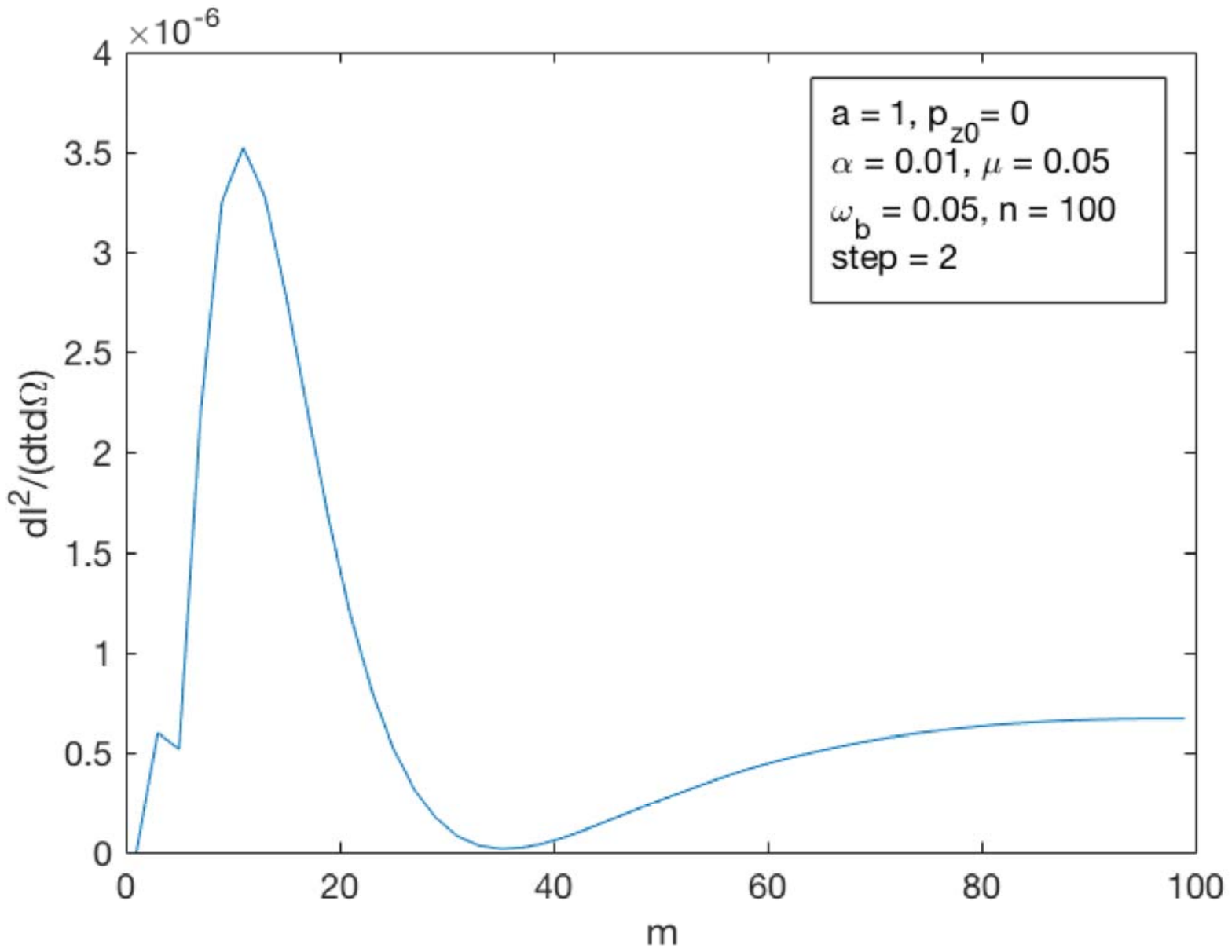}
\vskip -2.5cm
\caption{The backscatter spectrum of $m$th-order harmonic radiation. $a=1$, $p_{z0} = 0$, $\alpha = 0.01$, $\mu = 0.05$, $\omega_b = 0.05$, and $n = 100$. Odd harmonics are plotted from $m = 1$ to $100$.}
\end{figure}

In order to have a direct recognition of the intensity increased by our scheme presented here, we now compare some of the results in this paper to those in a well-known previous work \cite{FHe2003}. As what He \textit{et al}. have done, we denote the emission power in unit of erg/s as $p_m$ (see Eq.(5) of \cite{FHe2003}). Note that the normalizing intensity factor $e^2/4\pi^2 c$ is $0.69$ for $1\mathrm{\mu m}$ laser wavelength. For simplicity, we can approximate it as $\sim 1$. Two comparison examples will be given as the following. In the first example, we consider cases of Figs.2 (a) and 2(b), where the Thomson backscattering fundamental frequencies $\omega_1$ correspond to $\sim0.5$THz and $\sim2$THz. By matching these frequencies, the parameters of $\zeta$ of \cite{FHe2003} (which can be denoted as $\zeta_{prl}$ in order to be distinguished from that of this paper) are $\zeta_{prl}\sim 0.03$ and $\zeta_{prl}\sim 0.06$. In this case, the corresponding emission powers in \cite{FHe2003} are  $\sim 10^{-9}$ and $\sim 10^{-8}$ in unit of erg/s. On the other hand, our results indicate that, when a magnetic field $\sim 10^8$G is applied, the emission intensity is $\sim10^{-7}$ in Fig.2(a) for $\alpha=0$ and $\sim10^{-5}$ in Fig.2(b) for $\alpha=0.1$, which means that the presence of the magnetic field increases the intensity by 2 orders while the modulation of envelope further increases the intensity by another 2 orders. In the above example, our parameters correspond to the following system: laser intensity $10^{18}\mathrm{W/cm^2}$, wavelength $1\mathrm{\mu m}$ (laser frequency $\sim 10^{15}$Hz), magnetic field $10^8$G, and the initial electron rest energy $\sim0.5$ MeV; and the corresponding parameters of \cite{FHe2003} are the same except that there is no magnetic field and the initial electron energy is about $\sim(5-10)$MeV.
As the second example, we compare cases of Figs.3(a) and 3(b) to those in \cite{FHe2003}. Now the fundamental frequency $\omega_1$ is $\sim2.5$THz in Fig.3(a) and $\sim5$THZ in Fig.3(b). In this case, the emission intensity is $\sim 10^{-6}$ in Fig.3(a) and $\sim 10^{-4}$ in Fig.3(b) which are 2 to 3 orders higher than what were obtained in \cite{FHe2003}. The real physical parameters in this example are laser intensity $\sim 3\times10^{17}\mathrm{W/cm^2}$, wavelength $1\mathrm{\mu m}$ (laser frequency $\sim 10^{15}$Hz), magnetic field $\sim 5\times 10^7$G, and the initial electron rest energy $\sim0.5$ MeV. The corresponding parameters in \cite{FHe2003} are the same except that there is no magnetic field and the initial electrons energy is about $\sim(3-5)$MeV. It is worthy to point out that the laser fields of \cite{FHe2003} are linear but ours are circular. However, as shown in our previous work \cite{Jiang2017}, the emission intensities of the linear field and the circular field are within the same order (only 2-3 times different). In addition, the laser intensity of linear to circular is about 2 times different. Therefore, the comparisons mentioned above are reasonable in the sense of magnitude orders.

\section{Conclusion and Discussion}

In this paper, the Thomson backscattering spectra of an electron
moving in the combined cosine-enveloped laser and magnetic fields
have been studied. We have examined the effects of the
cosine-envelope, the cyclotron frequency $\omega_b$, the enveloping
coefficient $\alpha$, the circular frequency coefficient $\mu$,  the
laser intensity $a$, the constant of motion $\zeta$ and the initial
axial momentum $p_{z0}$ on the radiation spectra.

As demonstrated in our numerical examples, with the introduction of
the cosine-envelope, the radiation spectra exhibit complex and
striking phenomenon. High oscillations appear in the radiation
spectra, attributed to the strong nonlinear interactions. These
oscillations can be further analyzed when extracted into ARS curves.
We found that, for the same resonant parameter, a higher cyclotron
frequency will produce more ARS curves and, similar to the effects
of the enveloping coefficient, create an intense  radiation spectra
at higher harmonics. The circular frequency coefficient will shift
the peaks of the radiation spectra to higher harmonics, but it has a
negative correlation with the intensity of radiation spectra.
Furthermore, the laser intensity, the constant of motion, and the
initial axial momentum have an indirect correlation with the
radiation spectra intensity as proven in three of the four major
scaling laws of this study.

Analytically, we have derived and revealed four fundamental scaling
laws for the case of the cosine-enveloped laser field that are
upheld by the numerical results from this study. The scale
invariance and scaling law of the constant of motion is described as
$(d^2 I_m / dt d\omega) \propto \zeta^6$, and the scale invariance
and scaling law of the enveloping coefficient, the circular
frequency coefficient, and the modified cyclotron frequency is
described as $(d^2 I_m / dt d\omega)' = \rho^2 (d^2 I_m / dt d\omega)$. Neither of these laws have been
previously discovered in other studies.

In particular, the fourth law is crucial to the amplification and
tunability of the radiation spectra for further applications. In the
plots given by this paper, by solely applying this law, radiation
spectra intensities of $10^{-2}$ were obtained, which is remarkably
higher than the intensities found in previous papers. In addition,
radiation peaks can be intentionally tuned to both the right
intensity and the right frequency.

Therefore it is expected to find that if one choose appropriately
the laser and magnetic field parameters, the emitted spectral
broaden width may be controlled and minimized, and the peak
brightness of the emitted radiation can be increased by a existing
scale factor of  this study approximately. For example, in our last
example case, we successfully applied the fourth fundamental law to
produce high energy radiation in THz frequencies at an intensity of
about $3 \times 10^{-6}$, which is $10^{3}$ higher than what was
obtained before \cite{Jiang2017}. The spectral bandwidth reduction
of Thomson scattered light should be attributed the nonlinear
interferences arising from the pulsed nature of the laser
\cite{Umstadter2013}, which is also exhibited in our study here even
if by using a simple cosine-envelope laser modulation. The findings in this research are believed to have a large potential to greatly enhance radiation technology for a number of applications from imagining to remote sensing to communications.

\section*{Acknowledgements}

Much of the research was done by the first author under the mentorship of the second author. The first author is also grateful to C. Jiang, Z. Chen, and L. Zhao for some useful discussions. BSX is partially supported by the National Natural Science Foundation of China (NSFC) under Grant No. 11875007.

\begin{appendix}

\section{The determining of parameter $n$ related to the period solution of electrons}

From the solutions of electron about momenta and energy Eqs.(10-13), and also the positions Eqs(14-16), it is not difficult to see that there are three basic frequencies as
\begin{align}
\Omega_1=\mu+\alpha,\\
\Omega_2=\mu-\alpha,\\
\Omega_3=\omega_b,
\end{align}
and the three derivative difference frequencies among them as
\begin{align}
\Omega_4=\Omega_3-\Omega_1 = \omega_b-(\mu+\alpha),\\
\Omega_5=\Omega_3-\Omega_2 = \omega_b-(\mu-\alpha),\\
\Omega_6=\Omega_1-\Omega_2 = 2\alpha.
\end{align}
Therefore there exists an integer $n$ that satisfy the following algebraic equations, which make the solutions of momenta and positions are periodic except the net of displacement of $z$,
\begin{align}
n(\mu+\alpha)=l_1,\\
n(\mu-\alpha)=l_2,\\
n\omega_b=l_3,\\
\end{align}
where $l_i$ ($i=1,2,3$) are some integers. Obviously the $n\Omega_i=l_i$, where $i=4,5,6$, are automatically satisfied.

Now for simplicity of numerical calculation and in fact without losing generality let us set $l_4=1$ or $l_5=1$. For the former case
we have
\begin{align}
n\Omega_4=n/n_1=l_3-l_1=1,\\
n\Omega_5=n/n_2=l_3-l_2,\\
n\Omega_6=l_1-l_2.
\end{align}
This means that the $n_1/n_2=l$, where $l=l_3-l_2$ is an integer of $|l|\ge 1$, the equality or inequality corresponds to the case of $\alpha=0$ or $\alpha \neq 0$. So $n=n_1$ if we choose that
$$
\omega_b=\mu+\alpha+\frac{\mu+\alpha}{l_3-1}
$$
with some integer $l_3$. It is equivalent to that if our choosing about $\omega_b$ making $n \omega_b$ is an integer then there must exist an integer $l_3$ satisfying the above equations. Once $l_3$ exists, the other required integer conditions of $l_1=l_3-1, l_2=l_3-l, l_4=1, l_5=l$ and $l_6=l-1$ are all satisfied automatically. In particularly the involved periodic requirement of emission power intensity appeared in Eq.(22) are that there exist at least one set of the relative-prime integers among $1,l,l-1$. In fact there are two sets of relative-prime $1,l$ and $l,l-1$ when $l=2$, but complete three sets of relative-prime $1,l-1,l$ when $l>2$.

For the opposite case of $l_5=1$, where $n_2/n_1=l$ is an integer of $|l| \ge 1$, the similar analysis mentioned above can be performed and now $n=n_2$. Therefore we have get the condition that the parameter $n$ is determined by the combinational conditions of either $n_1/n_2$ or $n_2/n_1$ is an integer and $n = \max(n_1, n_2)$.
\end{appendix}

\end{document}